\documentclass[referee]{aa501}

\begin{document}

\title{Near infrared observations of the truncation of stellar disks}

\author{E. Florido \inst{1}, E. Battaner \inst{1}, A. Guijarro
\inst{1}, F. Garz\'on \inst{2} \and
J. Jim\'enez-Vicente \inst{3}}

\offprints{E. Battaner}

\institute{Dpto. F\'{\i}sica Te\'orica y del Cosmos, Universidad de Granada,
          Spain
          \and Instituto de Astrof\'{\i}sica de Canarias,
          V\'{\i}a L\'actea, s/n, La Laguna, Tenerife 
       \and
          Groningen Kapteyn Laboratorium, Groningen, Netherlands}
          	
\date{}
\authorrunning{E. Florido et al.}

\abstract{We present a first study of truncation of the stellar disks
of spiral galaxies in the near infrared. Observations of
NGC4013, NGC4217, NGC6504 and NGC5981 
were made with the CAIN NIR camera on the CST
in Tenerife. This wavelength range provides the best description of
the phenomenon, not only because extinction effects are minimized,
but also because the distribution of the old stellar population is
directly obtained. The four galaxies are edge-on and an inversion
method was developed to obtain the deprojected profiles. We did not
assume any model of the different galactic components. The
``truncation curve'', i.e. $T(R) = \mu(R)- \mu_{\rm{D}}(R)$, where $\mu$ is
the actual surface brightness in $\rm{mag} /\rm{arcsec}^2$ and
$\mu_{\rm{D}}$ the exponential disk surface brightness, has been obtained with
unprecedented precision. It is suggested that $T(R)$ is proportional
to $(R_{\rm{t}}-R)^{-1}$, where $R_{\rm{t}}$ is the truncation radius, i.e. the
radius beyond which no star is observed.
}

\maketitle

\section{Introduction}

At large radii the stellar density of disks decreases faster
than an exponential until reaching a cut-off or truncation radius
$R_{\rm{t}}$, where it vanishes. This morphological feature was discovered by
van der Kruit (1979) and later studied in more detail by van der Kruit
\& Searle (1981a, b \& 1982). Recently, this phenomenon has been
reconsidered by means of samples larger than the seven edge-on
galaxies observed by van der Kruit and Searle and by improved
observational techniques. Barteldrees \& Dettmar (1994), Pohlen,
Dettmar \& L$\ddot{u}$tticke (2000), Pohlen et al. (2000) and de Grijs
et al. (2001) have provided the basic information about truncations in
the optical range for external galaxies. Our Galaxy also presents a
truncation, although it is more difficult to observe (Habing 1988;
Robin et al. 1992; Ruphy et al. 1996; Freudenreich 1998). Porcel et al.
 (1997) found that the Milky Way cut-off
radius cannot be placed at distances larger than 15 \rm{Kpc}. 
Truncations of stellar disks have been reviewed by van der Kruit
(2000). Much remains to be done both from theory and observations to
understand this phenomenon.

A) Lack of theoretical explanation

The above studies have established the universality of the
phenomenon. Most galaxies, if not all, seem to have truncated stellar
disks, sensitivity limits alone are unable to explain this feature. This fact
emphasizes the theoretical importance of the topic. However,
 truncations constitute one of the most
important challenges in galactic dynamics. Though several hypotheses
have been considered, this phenomenon remains completely unexplained.

It was suggested by van der Kruit (2000) that stellar
truncation is accompanied by a significant drop in rotation velocity,
with NGC4013, NGC891 and NGC5907 being clear examples of this. If this
fact is confirmed, and actually takes place in most truncated disks, it
would mean that
there is a true decrease in the radial distribution of the total
density, i.e. the sum of both, the gas and the stellar densities. Theories
suggesting that stellar truncation is due to a cut-off of the star
formation rate beyond a certain radius should be reconsidered, as in
this case the total
gas plus star density would not present any discontinuity. The
confirmation of a drop in rotation velocity close to the stellar
truncation would pose serious difficulties for the most promising
hypothesis, maintained by Kennicutt (1989) and others, in which star
formation does not proceed when the gas density is lower than a
certain threshold value and would reject all theories in which stars
do not exist beyond the truncation radius, because they are not formed.

Another argument against the absence of star formation as the cause of
stellar truncation is that we do see star formation beyond the
truncation radius of the Milky Way. Molecular clouds are often
associated with HII regions, IRAS sources, H$_{\rm{2}}$O masers and other
objects characterizing the presence of the formation of high mass
stars (Mead et al. 1987; Mead et al. 1990; Brand \& Wouterloot 1994;
Rudolph et al. 1996; Williams \& McKee 1997; May et al. 1997; Kobayashi
\& Tokunaga 2000, and others). A high star formation rate is also
observed in other galaxies (Lequeux \& Guelin 1996). Wouterloot,
Brand \& Henkel (1988) and Ferguson et al. (1998) found the
important result that the amount of star formation per unit mass of
H$_{\rm{2}}$ at $R$ = 15 \rm{Kpc} is equal to that in the solar
neighbourhood. The ratio $N(HII) / \sigma(H_{\rm{2}})$ at $R$ = 15
\rm{Kpc} was found to be higher (by a factor 10/7) than in
the solar neighbourhood. Though some differences are found between the
outer galaxy molecular clouds and the inner ones at $\sim R_\odot$, they
have much in common, such as a similar star formation efficiency (Santos
et al. 2000) and kinetic temperature (Brand \& Wouterloot
1996). The similarities are more noticeable if we compare molecular
clouds at $R > R_{\rm{t}}$ (where $R_{\rm{t}}$ is the truncation
radius) and $R < R_{\rm{t}}$ but close to $R_{\rm{t}}$. This was done
by Brand \& Wouterloot (1991, 1994) and Wouterloot et al. (1993) with
their sample for 16\rm{Kpc}$<R<$ 20 \rm{Kpc} and the sample by Mead \& Kutner
(1988) for $R \sim$ 13 \rm{Kpc}. More information about
molecular clouds beyond the solar radius has been provided by Brand \&
Wouterloot (1995), Wouterloot \& Brand (1996) and Wouterloot et al. (1995,
1997). The range of masses and sizes are very
similar, and  hence the densities should be similar.
Cloud formation could be much more inefficient than at smaller $R$ 
(Brand \& Wouterloot 1991). There is indeed, a sharp decrease in
H$_{\rm{2}}$, but not that pronounced in HI gas, which might suggest that the
formation of molecular clouds out of HI gas is not as efficient beyond
some radius. Small unobservable clouds could have no star forming
capacity. However, the sudden step of the rotation curve mentioned by
van der Kruit (2000) remains unexplained.

A simplified but reasonable picture would then be: the amount of
molecular hydrogen and the number of clouds decrease; but the
density within a cloud remains more or less constant; therefore, if
there is a minimum H$_{\rm{2}}$ density for star formation it cannot
explain the truncation of the stellar disk.

Then the puzzling question is: if there is star
formation beyond $R_{\rm{t}}$, where are the stars? There are two
possible answers: a) Star
formation at these large radii is a recent or transient process, so
that stars have not been continuously filling this region. Suppose, for
instance, that the outer disk has been formed recently, because the
disk forms slowly and its radius increases over time. This
hypothesis is considered as a possibility by de Grijs et al.
(2000) and van der Kruit (2000) and has some theoretical
support from early works by Larson (1976) and Gunn (1982). Given our
present uncertainties about disk formation, though, this hypothesis is
rather speculative.
b) Stars, once born, then migrate away. This could be the case if stars and gas
have different dynamical behaviours, being subject to different
forces. Newly formed stars could be subject to other forces and
migrate from their birth place.

B) Observational problems.

There are two basic features in the surveys and analysis
carried out until now which are improved in this work. First, previous
observational studies have been made at optical wavelengths, and
therefore extinction introduces a severe limitation on the
interpretation of the results. Second, the analysis is usually based
upon a specific galaxy model with various components of which the
space distribution is specified by means of a number of free
parameters, which are determined by fitting the observations. However,
with this procedure, what is obtained is, in part,
what is assumed. Mathematical expressions are still insufficient for
many galaxy components.

This fact is specially problematic in the truncation region. It is
known that the truncation is not completely sharp, but rather starts
as a smooth deviation of the ``exponential'' disk (i.e. linear when
using $\mu$). A truncation curve $T(R)$ would quantify this smooth
deviation and can be defined precisely as $T(R) = \mu(R) -
\mu_{\rm{D}}(R)$ where $\mu(R)$ is the observed surface brightness in
$\rm{mag}/\rm{arcsec}^2$ and $\mu_{\rm{D}}(R)$ is the exponential
surface brightness extrapolated from the inner disk.
We know $T(R_{\rm{t}})= \infty$ in units of $\rm{mag}/\rm{arcsec}^2$,
where $R_{\rm{t}}$ is the truncation radius. Previous analysis have mainly
considered $R_{\rm{t}}$. Truncations are an
interesting object of study, as they could reveal the historical and dynamical
properties of a galaxy. However the whole
truncation curve, $T(R)$ also contains valuable information. It is
therefore worrying that the mathematical expression of $T(R)$ was
assumed rather than obtained as a chief objective.

To avoid the extinction deformation of the radial profiles, we have
observed in the near infrared, so we are mostly dealing with the old stellar
population. We present observations in J and Ks. Extinction in J is
more severe than it is in Ks. Therefore, conclusions obtained
from our measurements in Ks are more reliable. 

Complementary studies in other colours has been
addressed by the above cited texts. NIR CCD-like arrays already exist
some years, but the recent improvement of
two-dimension detectors and, in particular, that of 
CAIN, has made it possible to reach the truncation region.

To avoid model-dependent results, we
have used a numerical inversion method. Binney and Tremaine (1987)
describe another analytic method to carry out this deprojection, based
on the Abell integral. In our procedure, however, only two assumptions are
necessary: axisymmetry and negligible extinction. These two conditions are
by no means guaranteed in a disk galaxy but it should be taken into
account that axisymmetry is implicitly assumed in other
procedures. Moreover, we have two sides in an edge-on galaxy, which
are all very similar in our sample. Even if a non-axisymmetric
disk could exhibit two similar sides when a galaxy is seen edge-on,
this is rather improbable. Extinction is a problem when using methods based on
previous modeling, as dust often has a ring structure rather than
following an exponential law and the dust distribution must be risky
prescribed. Dealing with NIR observations and observing that our
$\mu(z)$-profiles do not show a secondary minimum produced by a dust
lane, inspires confidence in our method. Also, extinction is probably
no longer important in the peripheral truncation region.

With this small number of assumptions required, we obtain a
non-model-dependent deprojection.

\section{Observations and reduction}

The observations were carried out at the 1.5 m CST in the Teide Observatory,
Tenerife, with the NIR camera CAIN. This is a common user 2D NIR image camera
equipped with a $256^2$ NICMOS detector array. Two different plate
scales (0.4 and 1.0 \rm{arcsec}/\rm{pixel}) are selectable to obtain a
narrow or wide field image. We used the wide field optics, which has an
effective field of view of $4.3 \times 4.3$ \rm{arcmin}. The objects
were selected according to their projected size ($D_{\rm{25}}$) to fit
within that FOV. The detector control and read--out system was, at the
time of the observations, based on dedicated transputer design
electronics. In June 1999, the electronics was upgraded to a new
design based on San Diego State University controller, adapted in
house to the NIR, which provides better noise figures and
stability. The transputer controller exhibits several noise correlated
patterns which have to be removed during the reduction process by the
use of specifically designed software routines.

The observed galaxies were the edge-on galaxies NGC4013, NGC4217, NGC6504 and
NGC5981. They were observed in the period 13-19 April 
1999, as shown in Table 1. The basic physical parameters of these
galaxies are shown in Table 2.

\begin{table}
\caption{Observational parameters for the galaxies}
\begin{flushleft}
\begin{tabular}{|cccccc|}
  \hline
  Galaxy & Day & Passband & Exp. time & Seeing & 3$\sigma$ level  \\
         &      &         & object+sky(m) &    & \rm{mag}/\rm{arcsec}$^2$  \\
  \hline
  NGC4013 & 17,19 & Ks & 96 & 1.94  & 20.6 \\
          & 13,17 & J  & 80 & 1.80  & 22.0 \\
          & & & & & \\
  NGC4217 & 16 & Ks & 96 & 1.27  & 20.4 \\
          & 18 & J & 80 & 1.39 & 21.8 \\
          & & & & & \\
  NGC5981 & 20 & Ks & 64 & 1.45 & 19.9 \\
          & & & & & \\
  NGC6504 & 16 & Ks & 88 & 1.27 & 20.4\\
          & 15 & J & 64 & 2.96 & 21.5\\
          & 18 & H & 64 & 1.57 & 20.5\\
  \hline
\end{tabular}
\end{flushleft}
\end{table}

\begin{table}
\caption{Physical parameters for the observed galaxies obtained
from LEDA database (http://leda.univ-lyon1.fr): RA and DEC are the right ascension and declination in 2000,
PA is the position angle, $d$ the distance and mabs the
absolute B-magnitude}
\begin{flushleft}  
\begin{tabular}{|c|c|c|c|c|c|c|}
  \hline
  Galaxy & RA & Dec & Type & PA &d & $m_{abs}$  \\
         &$^{(h \; m \; s)}$ & $^{(o \; ' \; '')}$ & &$^{o}$ & (\rm{Mpc}) &  \\
  \hline
  NGC4013 & 11 58 31.5 & 43 56 51 & Sb & 66 & 11.16 & -19.47  \\
          & & & & & & \\
  NGC4217 & 12 15 50.8 & 47 05 30.8 & Sb & 50 & 13.64 & -19.96\\
           & & & & & & \\
  NGC5981 & 15 37 53.5 & 59 23 28.7 & Sbc & 140 & 30.18 & -20.34\\
          & & & & & &  \\
  NGC6504 & 17 56 5.7 & 33 12 31.7 & Sbc & 94 & 63.18 & -22.28 \\
  \hline 
\end{tabular}
\end{flushleft}
\end{table}

In order to correct for the bright and rapidly varying NIR
sky background,
the telescope was alternatively pointed to six fields in the sky in
the order $O_1S_1S_2O_2O_1S_3S_4O_2$, where the $S's$ are background
and the $O's$ contain the galaxy. The $O_2$ field was offset with
respect to $O_1$, 15'' N and 15'' E. $S_1$ was 600'' W
from $O_1$; $S_2$, 900'' W; $S_3$, 600'' E from
$O_2$; $S_4$, 900'' W. Each exposure lasted about 2 minutes.

It was very important to perform good flat fielding, sky subtracting
and mosaicing. We used the data reduction package developed by
R. Peletier, REDUCE, within IRAF, which is specially suitable for data
with a large sky background. We took object
images, bias frames at the beginning and/or at the end of the night,
dark frames for the two exposure times used (10 and 30 sec) and
flatfields to calibrate the sensitivity of the array. We took bright
and dark flatfields for each filter with the same integration time; 
these were then combined and subtracted to remove the effects of dark
current, telescope and dome.

The calibration was done by using the UKIRT Faint Standard Stars
(Casali \& Hawarden 1992)
fs18, fs23, fs24, fs27 and fs28. We took 4 blocks of 15 images each, for
every filter and for every star, at least three times per night, for
different air masses. 
After calibration the isophote contour maps for
the four observed galaxies (see Fig. 1) were obtained by means of IRAF Newcont.

\begin{figure}
  \caption[]{Contour maps for the observed galaxies: The
interval between isophotes is 0.5 magnitudes/\rm{arcsec}$^2$ in all
maps. The lower value is 15 magnitudes/\rm{arcsec}$^2$ for NGC5981, NGC4013
in Ks, NGC6504 in Ks and NGC6504 in H; 15.5 mag/\rm{arcsec}$^2$ for NGC6504
in J and NGC4217 in Ks and 16 for NGC4013 in J. East is at bottom and
North on right}
\end{figure}

\section{The deprojection method}

We divide the disk into rings with constant $\Delta R$ and assume a
constant emissivity within a ring, i.e. a constant emission per unit
volume in the direction of the observer $l_i$. (See Fig. 2).
>From the edge-on surface brightness, $I$, we must deduce $l_i$, taking
into account that many rings contribute to the integral $I$, being the
contribution of each ring weighted by a different area. Once $l_i$
is obtained we must integrate in the vertical direction to determine
what would be seen if the galaxy were face-on. First, in the
equation
\begin{equation}
  I_k = 2 \Sigma_i A_{ki} l_i
\end{equation}
we must obtain $l_i$ through an inverse method. The index $k$ denotes
the different pixels under observation for $z=0$. The index $i$
denotes the different rings in which the galactic plane is
divided. $A_{ki}$ are the areas shown in Fig. 2.  

\begin{figure}
  \caption[]{Scheme for the inversion method. Subindex $i$ denotes the
  disk rings. Subindex $k$ denotes radial distances on the edge-on
  observed galaxies}
\end{figure}

If $k >0$
\begin{equation}
 \begin{array}{rl}
  A_{ki} = & \int_{(k-1/2)\Delta}^{(k+1/2)\Delta} \left( \left( i +
  1/2\right)^2 \Delta^2 - x^2 \right)^{1/2} dx \\
  & \\
  & - \underline{\int_{(k-1/2)\Delta}^{(k+1/2)\Delta} \left( \left( i -
  1/2\right)^2 \Delta^2 - x^2 \right)^{1/2} dx}
 \end{array}
\end{equation}

where $x$ is an integration variable, $\Delta$ is the pixel size.

The second integral must not be calculated for $i = k$; this is the
meaning of the symbol $\underline{\hskip 2cm}$ placed below the
integral. In all cases, we must have $i \geq k$.

If $k=0$
\begin{equation}
 \begin{array}{rl}
  A_{0i} = & 2 \int_0^{\Delta/2} \left( \left( i +
  1/2\right)^2 \Delta^2 - x^2 \right)^{1/2} dx \\
  & \\
  &- \underline{2\int_0^{\Delta/2} \left( \left( i -
  1/2\right)^2 \Delta^2 - x^2 \right)^{1/2} dx}
 \end{array}
\end{equation}
The second integral must not be calculated for $i=0$.

As an example,
\begin{equation}
 \begin{array}{rl}
  A_{00} &= 2\int_0^{\Delta/2} \left( {{\Delta^2} \over 4} - x^2 \right)^{1/2} dx =
          \left[ x \left({{\Delta^2} \over 4} - x^2 \right)^{1/2}  +
          {{\Delta^2} \over 4} 
          \sin^{-1}{x \over {\Delta/2}} \right]_0^{\Delta/2} \\
	  &  \\
          & =  {{\Delta^2} \over 4} \sin^{-1}{{\Delta/2} \over {\Delta
          /2}}        
          = {{\Delta^2} \over 4} \sin^{-1}{1} = {{\Delta^2} \over 4}
          {\pi \over 2}
 \end{array}
\end{equation}

In fact, this is a semicircle of radius $\Delta/2$, with area ${1
\over {2 \pi}} (\Delta/2)^2$.

In general, to calculate the above integrals we  take into
account that: 
\begin{equation}
  \int \left( a^2-x^2 \right)^{1/2} dx = {1 \over 2} \left[ x \left(
  a^2 - x^2 \right)^{1/2} + a^2 \sin^{-1}{x \over {\mid a \mid}}
  \right]
\end{equation}
We thus obtain
\begin{equation}
 \begin{array}{rl}
  2A_{ki} =& \Delta^2 \left( \left( k+{1 \over 2} \right) \left( \left(
  i + {1 \over 2} \right)^2 - \left( k + {1 \over 2} \right)^2
  \right)^{1/2} - 
      \left( k-{1 \over 2} \right) \left( \left(
  i + {1 \over 2} \right)^2 - \left( k - {1 \over 2} \right)^2
  \right)^{1/2} \right. \\
   &            \\
   & - \underline{\left( k+{1 \over 2} \right) \left( \left(
  i - {1 \over 2} \right)^2 - \left( k + {1 \over 2} \right)^2
  \right)^{1/2}} +
      \underline{\left( k-{1 \over 2} \right) \left( \left(
  i - {1 \over 2} \right)^2 - \left( k - {1 \over 2} \right)^2
  \right)^{1/2}} \\
  &              \\
  & + \left(i + {1 \over 2} \right)^2 \sin^{-1}{{k+1/2} \over {i+1/2}} -
  \left(i + {1 \over 2} \right)^2 \sin^{-1}{{k-1/2} \over {i+1/2}} \\  
  &  \\
  &  \left. \rule{0cm}{0.5cm} 
   -\underline{ \left(i - {1 \over 2} \right)^2 \sin^{-1}{{k+1/2} \over
  {i-1/2}}} +    
  \underline{\left(i - {1 \over 2} \right)^2 \sin^{-1}{{k-1/2} \over
  {i-1/2}}}  \right)
 \end{array}
\end{equation}

Once the areas $A_{ki}$ are calculated we need only obtain the inverse
matrix. But there is a much simpler procedure, beginning with the
maximum $k$ pixel ($k = N$) and going backwards until $k=0$.

For $k = N$
\begin{equation}
  I_{\rm{N}} = 2 A_{\rm{NN}} l_{\rm{N}}
\end{equation}
\begin{equation}
  l_{\rm{N}} = I_{\rm{N}}/2A_{\rm{NN}}
\end{equation}

In general
\begin{equation}
  l_k = {{I_k - 2\Sigma_{i=k+1}^N A_{ki} l_i} \over {2 A_{kk}}}
\end{equation}
because to calculate $l_k$ we already know all $l_m$ for $m > k$.

If we repeat this procedure for each $z$ we obtain a matrix
$l_{iz}$. The deprojection simply implies the integration of $l_{iz}$
over $z$, and hence the surface brightness if the galaxy were observed
face-on, i.e. for the deprojected galaxy, $I_i$ would simply be
\begin{equation}
  I_i = \Sigma_z l_{iz}
\end{equation}

The determination of $N$, the last measurable point at which the
galaxy signal and the sky background merge, is important. Several
initial trials indicated that the deprojection was not affected by the
choice of $N$, with the exception of the truncation region. Even in
this case, the effect was 
only significant when the value of $N$ was too large. We have adopted
the 3$\sigma$-criterion.

\subsection{The errors introduced by the inversion method}

Two sources of errors arise from the method itself, one of which is due
to the choice of the last point considered as the transition between
galaxy and background sky. As the value of the deprojected surface
brightness of the more external points partly decides the values at a
given position, the choice of the starting point is important. After
some trials it was found that the standard 3$\sigma$-criterion
was satisfactory and only unrealistic (much higher or lower)
starting points produced a significantly different truncation curve.

Another source of errors arises from the formula involved, which
implies a propagation of errors. The
deprojected profile is slightly noisier than the observed one. Doubts could
arise as to whether an error at a point, due for instance to a foreground
star or just due to an error inherent in the instruments, could have
a large influence at points far from the star. To evaluate the typical
errors introduced by this effect, let us consider equation (9).

Suppose the individual errors in surface brightness, due to
observation and instrumental errors, are $\Delta I_k \equiv \Delta I$, all
approximately equal (around 2$\sigma$). With $A$ being a typical value of
$A_{ik}$ we would have, approximately
\begin{equation}
 \begin{array}{rl}
  \Delta^2 l_k & = {1 \over {(2A)^2}} \Delta^2 I_k +
     \left( {{2A} \over {2A}} \right)^2 \sum_{i=k+1}^N \Delta^2 l_i \\
 & \\
 & = {1 \over {(2A)^2}} \Delta^2 I_k + \left( {1 \over {(2A)^2}}
  \Delta^2 I_{k+1} + \dots \right) =
  {1 \over {(2A)^2}} (N-k) \Delta^2 I
 \end{array}
\end{equation}
Therefore
\begin{equation}
  \Delta l_k \sim {{\sqrt{N-k}} \over {2A}} \Delta I
\end{equation}
Integrating in $z$, to obtain $I_{d,k}$ would therefore introduce an
error of
\begin{equation}
  \Delta I_{d,k} = \sqrt{{\cal N}_k} {{\sqrt{N-k}} \over {2A}} \Delta
  I
\end{equation}
where ${\cal N}_k$ is the number of layers in $z$, to be considered in
this sum. Hence
\begin{equation}
  \Delta \mu = (2.5 \log{e}) {{\sqrt{{\cal N}_k} \sqrt{N-k}} \over {2A}} {{\Delta
  I} \over I} \approx {{\sqrt{{\cal N}_k} \sqrt{N-k}} \over {2A}} {{\Delta
  I} \over I}
\end{equation}

For a point in the inner part of the galaxy, let us take as typical
values ${\cal N}_k \sim 20$; $N-k \sim 20$, $A \sim 1$, $I \sim 300
\sigma$, $\Delta I \sim \sigma$. Then, $\Delta \mu$ is obtained to be
of the order of 0.03 magnitudes.

For a point in the truncation region, ${\cal N}_k \sim 4$; $N-k \sim 4$, $A \sim 5$, $I \sim 10
\sigma$, $\Delta I \sim \sigma$. Then, $\Delta \mu \sim 0.04$ \rm{mag}.

For the last adopted point, ${\cal N}_k \sim 1$; $N-k \sim 1$, $A \sim 10$, $I \sim 3
\sigma$, $\Delta I \sim \sigma$, we again obtain 0.03 \rm{mag}.

We conclude that errors due to the method are of the order of 0.03 \rm{mag}
and that this error is constant throughout the disk. At the
rim, the large relative errors in brightness are compensated by a
larger typical area, $A$, as well as by the small numbers of
$z$-layers available. Errors in the method itself are therefore less
than 0.1 \rm{mag}.

\section{Results}

In order to apply this inversion method, we first rotated all
galaxies. The observed (projected) radial profiles were
obtained for the interval $z = 0 \pm \Delta z$, where $\Delta z$ was
small ($\approx$ 10-30 \rm{arcsec}), and different for each galaxy. We
used data only
up to the radius where the surface brightness was equal to $3
\sigma$. Particular care was taken when relatively bright stars were
at the rim of the galaxy.

We then applied the model for each colour, each side and each galaxy. 
The quantity $l_{iz}$ was obtained, and then the
intensity $I_i$. Results are shown in Fig. 3.

\begin{figure}
  \caption[]{Profiles obtained with the inversion method. Negative
  values in $R$ axis denote eastern side of the galaxies. Horizontal
  bars indicate the region in which the radial scale length was calculated}
\end{figure}

The bulge, the quasi-exponential disk and in some cases the ring, are
clearly visible. What is remarkable is the truncation, which is
readily observed. The truncation of all galaxies is defined in such detail
that the truncation region can be studied with unprecedented precision. 

Note that a large portion of the curves in Fig. 3 corresponds to
points below or far below the $3\sigma$-level given in Table 1. The
$3\sigma$ criterion was applied to the ``observed'' points, but the
profiles in Fig. 3 show ``calculated'' points. Due to the deprojection
method we are deducing values of the deprojected surface brightness
that would be unobservable if the galaxy were face-on, and which are
therefore below the observational $3\sigma$-level. This fact implies
that the truncation phenomenon is undetectable in face-on galaxies, at
least with the noise level of our observations. Also note that the
$3\sigma$ level is within the R-range used to fit the exponential
disks. But again we must take into account the difference between
observed and deprojected points. Observed surface brightness should be
obtained as an integration of deprojected surface brightness, and are
therefore much higher.

\subsection{The truncation curve}

First, we obtained the radial  scale length by excluding points belonging
either to the bulge or to the truncation region, or even some parts of the
galaxy where the radial variation was not a clear exponential
function (see the horizontal lines in Fig. 3). Assuming a law of
the type $I=I_0 e^{-R/R_{\rm{d}}}$ we obtained
\begin{equation}
  \mu = \mu_0 + {{1.086}\over {R_{\rm{d}}}} R
\end{equation}
where $\mu$ and $\mu_0$ are magnitudes per squared arc-second, $\mu_0=\mu(R=0)$,
$R$ is the galactocentric radius and $R_{\rm{d}}$ the radial scale
length. Results are summarized in Table 3. Our
results can be compared with those obtained by van der 
Kruit \& Searle (1982) for those galaxies present in both their and
our sample. They found $R_{\rm{d}}$=49.45'' versus our value of
48.38'' in J and 37.83'' in Ks for NGC4217. In the case of NGC4013
they obtained 40.34'' whereas our data give 40.06'' in J and 33.52'' in Ks. A
change in scalelength between the visual and NIR means that the colour
of the galaxy changes with R, either because of a true variation of
the stellar population or to a variation due to extinction.

\begin{table}
 \caption{Adjusted parameters for the disk. }
\begin{tabular}{|llcccc|}
  \hline
  Galaxy & Passband & Side & $R_{\rm{d}}$    & $R_{\rm{d}}$ & $\mu_{\rm{o}}$ \\
         &          &      & (\rm{arcsec}) & (\rm{Kpc}) & (\rm{mag}/\rm{arcsec}$^2$) \\
  \hline
  NGC4013&  J       &  NE   &  48   &  2.58 & 18.98 \\
         &          &  SW   &  32   &  1.75 & 18.12 \\
         &  Ks      &  NE  &  37   &  1.99 & 17.82 \\
         &          &  SW   &  30   &  1.63 & 17.38 \\
         &          &      &          &       &       \\
  NGC4217&  J       &  NE   &  53   &  3.51 & 18.93 \\
         &          &  SW   &  44   &  2.89 & 18.67 \\
         &  Ks      &  NE   &  35   &  2.35 & 17.28 \\
         &          &  SW   &  40   &  2.65 & 17.44 \\
         &          &      &          &       &       \\
  NGC5981&  Ks      &  SE   &  21   &  3.10 & 17.86 \\
         &          &  NW   &  24   &  3.45 & 18.01 \\
         &          &      &          &       &       \\
  NGC6504&  J       &  SE   &  25   &  7.54 & 19.07 \\
         &          &  NW   &  54   &  16.39& 19.89 \\
         &  Ks      &  SE   &  29   &  8.99 & 18.51 \\
         &          &  NW   &  38   & 11.57 & 18.83 \\
         &  H       &  SE   &  32   &  9.91 & 19.02 \\
         &          &  NW   &  32   &  9.75 & 18.99 \\
\hline
\end{tabular}
\end{table}

In the definition of the truncation curve
\begin{equation}
  T(R) = \mu(R) - \mu_{\rm{D}}(R)
\end{equation}
$\mu_{\rm{D}}(R)$ is equal to $\mu_0 + (1.086/R_{\rm{d}})R$, i.e. the function
$\mu(R)$ if it were a pure exponential. Plots of $T(R)$, the truncation
curve, are the main objectives of this research, and are given in
Fig. 4. Clearly, points of $T(R)$ in the bulge region are meaningless.

\begin{figure}
  \caption[]{Truncation curves. In the case
  NGC4013, SW side, Ks filter, the fitted profile has been included,
  as an example}
\end{figure}

In order to fit these curves, we have considered the
three-parameter function
\begin{equation}
  T= {a \over {(R_{\rm{t}} - R)^n}}
\end{equation}
where $a$ and $n$ are fitting parameters, as well as $R_{\rm{t}}$, the
truncation radius, because for $R=R_{\rm{t}}$ we obtain $T= \infty$, so there
is a complete truncation for $R=R_{\rm{t}}$. This function has been selected
because $T$ can be very small for the large inner range where $R \ll
R_{\rm{t}}$, but becomes larger for $R$ close to $R_{\rm{t}}$. Clearly
this function cannot be applied for $R>R_{\rm{t}}$.

The fit is outlined in table 4. In the particular case of the NW side
of NGC6504, a bright field star is located in the truncation region. We
have made different calculations with and without the star, starting before or beyond it, but the results did not differ
significantly. Nevertheless, our plot for the NW side of NGC6504
should be interpreted with extreme caution.

\begin{table}
 \caption{Fit parameters for the truncation curve}
 \begin{flushleft}
\begin{tabular}{|llcccccc|}
 \hline
  Galaxy & Passband & Side & $R_{\rm{t}}$    & $R_{\rm{t}}$ & $a$   & $n$ & $R_{\rm{t}}/R_{\rm{d}}$ \\
         &          &      & (\rm{arcsec}) & (\rm{Kpc}) &       &   &  \\
  \hline
  NGC4013&  J       &  NE   &  160     &  8.65 &  2.43 & 1.10 & 3.34\\
         &          &  SW   &  155     &  8.38 &  3.56 & 0.77 & 4.79\\
         &  Ks      &  NE   &  147     &  7.95 &  5.64 & 1.04 & 3.98\\
         &          &  SW   &  140     &  7.57 &281.46 & 1.90 & 4.64\\
         &          &      &          &       &       &      &     \\
  NGC4217&  J       &  NE   &  170     & 11.24 & 11.59 & 1.15 & 3.20\\
         &          &  SW   &  165     & 10.91 & 12.30 & 1.12 & 3.78\\
         &  Ks      &  NE   &  155     & 10.25 &  7.10 & 1.00 & 4.36\\
         &          &  SW   &  148     &  9.79 &  1.07 & 0.59 & 3.69\\
         &          &      &          &       &       &      &     \\ 
  NGC5981&  Ks      &  SE   &  73      & 10.68 & 1.27  & 0.68 & 3.44\\
         &          &  NW   &  90      & 13.17 & 55.70 & 1.62 & 3.81\\
         &          &      &          &       &       &      &     \\
  NGC6504&  J       &  SE   &  75      & 22.97 &  1.15 & 0.55 & 3.05\\
         &          &  NW   &  70      & 21.44 & 665   & 2.57 & 1.29\\
         &  Ks      &  SE   &  75      & 22.97 & 28.79 & 1.48 & 2.56\\
         &          &  NW   &  75      & 22.97 & 68.72 & 1.73 & 1.98\\
         &  H       &  SE   &  75      & 22.97 & 10.48 & 1.14 & 2.32\\
         &          &  NW   &  70      & 21.44 & 16.94 & 1.43 & 2.20\\
 \hline
\end{tabular}
\end{flushleft}
\end{table}

Values of $a$ have a large scatter. This parameter is related to the brightness close to the cutoff radius, compared with the 3$\sigma$ level. More attention should be paid to the
value of the parameter $n$, as it is an exponent, therefore
characterizing the function by which the stellar population abruptly
ceases. We obtain a mean value of 1.24 $\sim$ 5/4, including all
filters, sides and galaxies. The mean value for the Ks
filter is probably more representative as it is more free from extinction
distortions and more accurately accounts for the stellar
population. But in this case the result is remarkably similar: 1.25
with a r.m.s. of about 0.4. The closer integer (unity) fully lies in
the range of acceptable values.

The average truncation radii show a large dispersion for the different
galaxies. For the average Ks values we obtain: NGC4013, $R_{\rm{t}}=7.76
\rm{Kpc}$; NGC4217, $R_{\rm{t}}=10.02 \rm{Kpc}$; NGC6504, $R_{\rm{t}}=
22.97 \rm{Kpc}$; NGC5981, $R_{\rm{t}}= 12.42 \rm{Kpc}$. Probably the
truncation would be sharper when observed with a better
seeing. However no relation was found between derived truncation
parameters and seeing, and therefore this effect does not
significantly modify our results.

\section{Discussion}

A great deal of attention has been paid previously to the coefficient
$R_{\rm{t}}/R_{\rm{d}}$. In the early work by van der Kruit \& Searle
(1982) it was estimated to be about $4.2\pm0.5$. De Grijs et al. (2000) also
give values ($4.3\pm0.3$, $3.8\pm0.5$, $4.5\pm0.4$ and $2.4\pm0.3$)
for the four galaxies in their sample. Pohlen et al. (2000a, b)
found much lower values, around $2.9\pm0.7$ and Barteldrees \&
Dettmar (1994) give a mean value of $3.7\pm1.0$. In this work,
where extinction and inclination effects have almost been eliminated, we
also deduce lower values than van der Kruit \& Searle (1982),
with 3.2 being the mean value for all the profiles considered. If we
limit ourselves to the values for the longest wavelength, i.e. Ks, we
obtain for the four galaxies: NGC4013, 4.31; NGC4217, 4.03; NGC6504,
2.27; NGC5981, 3.62 (mean value of the two sides of the
galaxy). The mean value for Ks is $3.6\pm0.8$. Given the high
$\sigma$, our value is compatible with all previous results. Clearly,
a 4-galaxy sample is too small, and more information including more
types of galaxies is necessary. But the data for our galaxies fit very
well in Fig. 2 in Pohlen et al. (2000). So, we agree with their
statement that {\it large disks with regard to their scalelengths
$R_{\rm{d}}$ are short in terms of their cut-off radii}.

Because the extinction was assumed to be negligible in our NIR
data, we cannot draw any conclusions about it. But, we consider that
the Ks data give the most reliable results.

In general, extinction would produce an apparent higher value of the
radial scale length, while $R_{\rm{t}}$ would be less affected as
truncation takes place in regions where extinction is less
important. Therefore it is believed that the optical values for
$R_{\rm{t}}/R_{\rm{d}}$ are underestimated.

The difference would not be excessive, as we obtain lower but
comparable values for the two galaxies we have in common with van der
Kruit and Searle.
The slope of the points in Fig. 6 gives the $R_{\rm{t}}/R_{\rm{d}}$ ratio. In
Fig. 6 we observe a sharp correlation between $R_{\rm{t}}$ and the
absolute magnitude of the galaxies. The brighter the galaxy, the
larger the disk is found to be. This is not unexpected, but the
correlation seems to be tight.
Fig. 7 shows that $n$ is not clearly dependent on a typical
parameter of a spiral, such as $R_{\rm{d}}$.

\begin{figure}
  \caption[]{$R_{\rm{t}}$ versus $R_{\rm{d}}$. Squares, Ks filter; triangles, J-filter and crosses, H-filter}
\end{figure}

\begin{figure}
  \caption[]{Absolute M-magnitude versus $R_{\rm{t}}$. Symbols as in Fig. 5}
\end{figure}

\begin{figure}
  \caption[]{$n$ versus $R_{\rm{d}}$. Symbols as in Fig. 5}  
\end{figure}

We find some colour dependence in the truncation radii. This is quite
clear for NGC4217 (151'' in Ks, 167'' in J; this tendency is confirmed
by its value in the optical given by van der Kruit \& Searle (1982) of 202'').
It is also clear for NGC4013 (143'' in Ks, 157'' in J; this tendency
is confirmed by its value in the optical of 165'' given by van der Kruit and
Searle). However, NGC6504 does not
show this colour dependence, $R_{\rm{t}}$ being nearly independent of
wavelength. We have only one filter for NGC5981. We suggest that
$R_{\rm{t}}$ could be lower for the older population, but this
conclusion is not at all firm. This fact could indicate that the
stellar disk has grown, but the observational evidence is scarce to
ascertain any interpretation of this colour dependence.

The coefficient $n$ has no clear relation with other basic parameters of
the galaxies. In particular no relation with colour was found. We
have obtained 1.25, with the simplest possible value (unity)
compatible within the error range. The simplest truncation curve 
compatible with our data would therefore be 
\begin{equation}
  T \propto {1 \over {R_{\rm{t}} -R}}
\end{equation}

This is an important result. As the truncation is not sharp (e.g. de
Grijs et al. 2000) the shape of the truncation curve should be
investigated by theoretical methods, and compared with
eq. 18. The theoretical prediction of this truncation curve would be
more restrictive than that of any other observable parameters.

Asymmetries have been observed but these are not large. The parameter
\begin{equation}
  2 {{|(R_{\rm{t}}(E) - R_{\rm{t}}(W))|} \over {R_{\rm{t}}(E) + R_{\rm{t}}(W)}}
\end{equation}
(where E and W denote the two sides of the galaxies)
characterizing the relative degree of asymmetry was only 0.05, 0.05,
0.00 and 0.21, for NGC4013, NGC4217, NGC6504 and NGC5981 respectively,
if we only take into account the more confident values of Ks.

A potentially very restrictive fact suggested by van der Kruit (2000), that
the rotation curve has a decreasing step just after the truncation,
can neither be rejected nor confirmed as the rotation curve of
NGC4013 (Verheijen 1997) shows this step 4 \rm{Kpc} after
$R_{\rm{t}}$, but it is  apparently not present for NGC4217 (see also
Verheijen 1997).

\section{Conclusions}

The stellar disk is less extended than the gaseous disk and this is
not the result of sensitivity limits of optical telescopes: there is
a physical mechanism that produces a relatively sharp truncation. To
investigate this mechanism, the advantages of observing in the near infrared
are obvious as we are looking for a phenomenon concerning the stars,
and because the results are not as affected by extinction problems,
whose radial distribution is not ``a priori'' known. The advantages of an
inversion method not prescribing ad-hoc mathematical models for the
different galactic components have been shown. In particular,
the use of a model with several free parameters for the truncation
region prevents us from obtaining the function we seek, i.e. the
detailed way in which the stellar component leaves the exponential
distribution and becomes completely truncated at $R_{\rm{t}}$.

To investigate the mechanism responsible for the truncation of the
stellar disk, more observational effort is needed, especially in the
near infrared. Also, more theoretical work is needed as many of the
hypotheses advanced up to now fail to explain 
firm constraints that have been established by observations.

The possible
relation between $R_{\rm{t}}$ and warps, which was considered
by van der Kruit (2000) is not clear: for NGC4013  and NGC6504, we can
see in Florido et al. (1991) that the warps begin at a radius that is
much smaller than $R_{\rm{t}}$.

Two important conclusions can be extracted from this work:

-The coefficient $R_{\rm{t}}/R_{\rm{d}}$ has been found to be lower
 than in some previous works, being for Ks $3.6\pm0.8$, at least for the four
 galaxies observed.

-We suggest that a colour dependence of the truncation radius exists:
 truncation seems to take place at lower radii for larger
 wavelengths, but this was only for 2/3 of the small sample. More
 observations are needed.

We present plots of the truncation curve $T(R)$, which is one of the
main objectives of this work. These plots are more informative than
the truncation radius itself. Nevertheless, we
have tried to fit these $T(R)$ curves. A reasonable candidate formula
is of the type $T(R) \propto (R_{\rm{t}}-R)^{-n}$, where the exponent $n$ has
been found to be $1.25\pm0.4$. This suggest $n=5/4$, with unity being
a value within the errors. A simple suggestion is therefore
\begin{equation}
  T(R) \propto 1 / (R_{\rm{t}}-R)
\end{equation}

\begin{acknowledgements}
We are thankful to Dr. Peletier for providing his very efficient
reduction software REDUCE and for many helpful discussions during the
period of this work.

The CST is operated on the island of  Tenerife by the Instituto de
Astrof\'{\i}sica de Canarias in the Spanish Observatorio del Teide of the
Instituto de Astrof\'{\i}sica de Canarias.

This paper has been supported by the ``Plan Andaluz de Investigacion''
(FQM-108) and by the ``Secretar\'{i}a de Estado de Pol\'{i}tica
Cient\'{i}fica y Tecnol\'ogica'' (AYA2000-1574).
\end{acknowledgements}

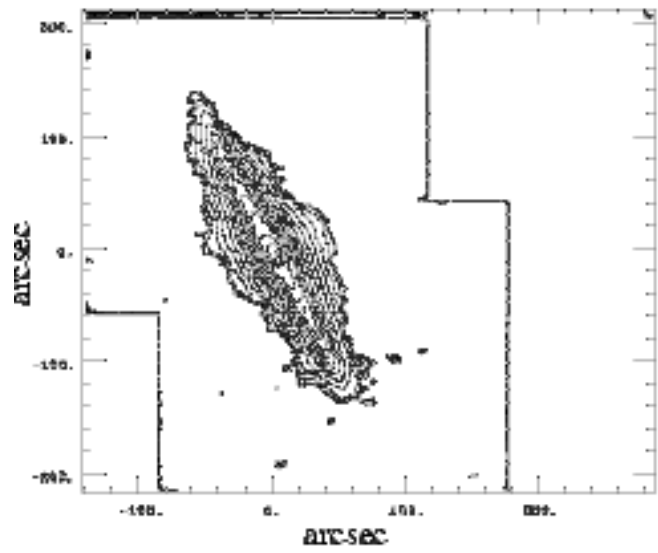
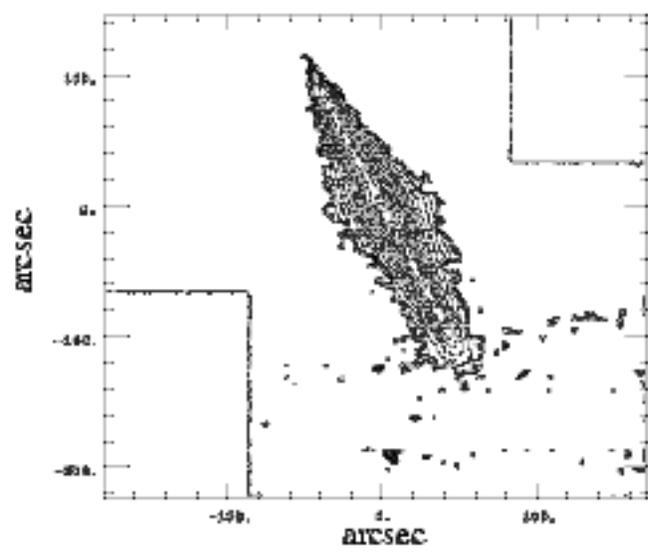
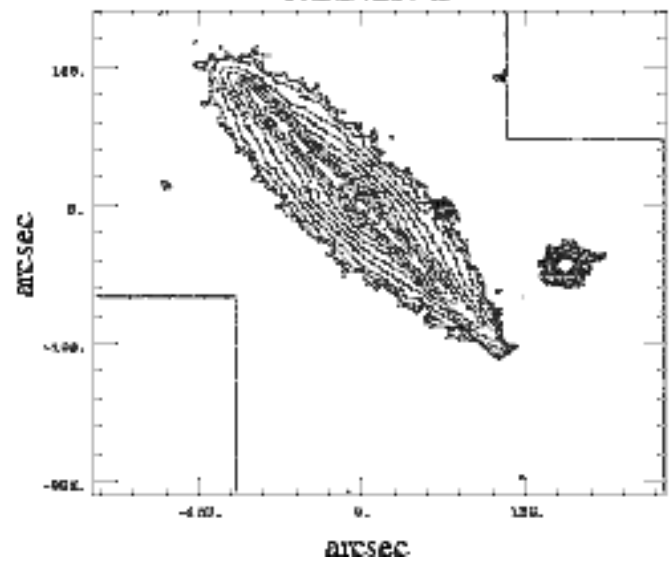
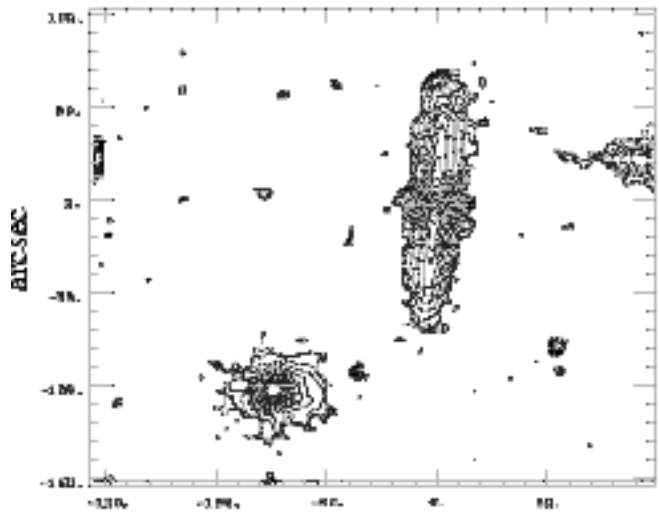
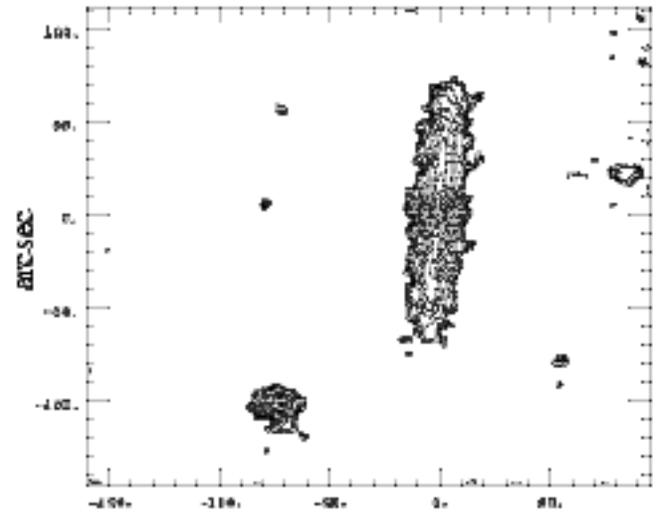
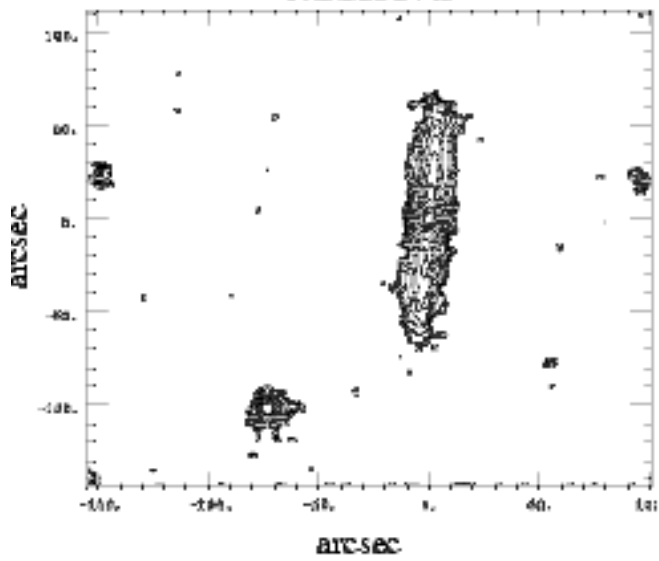
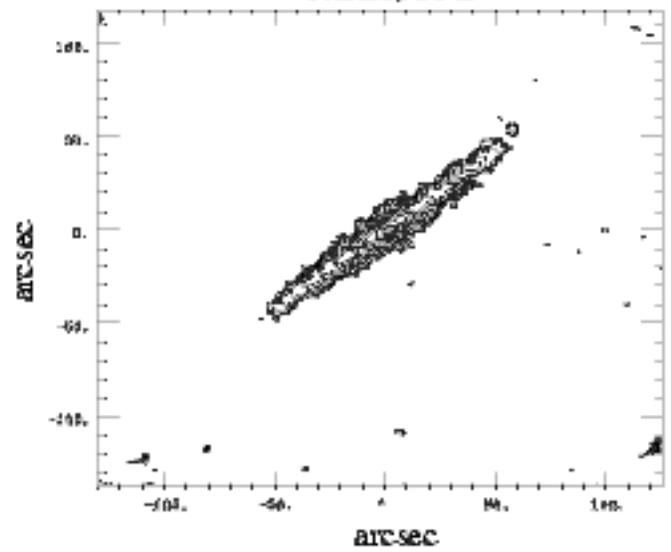
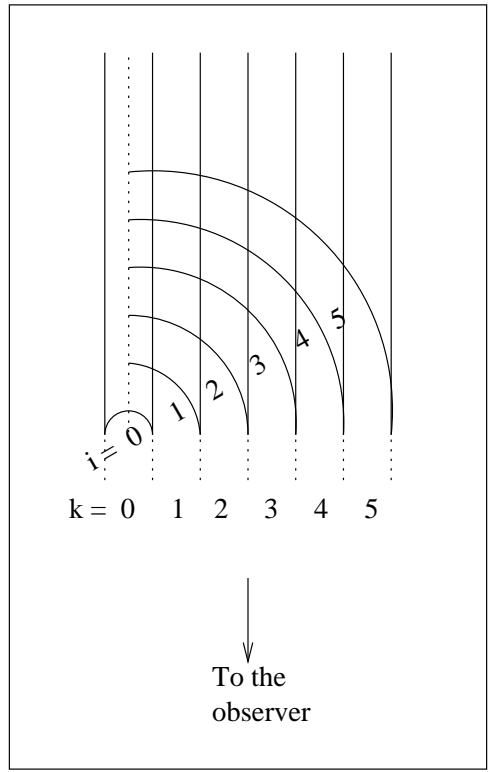
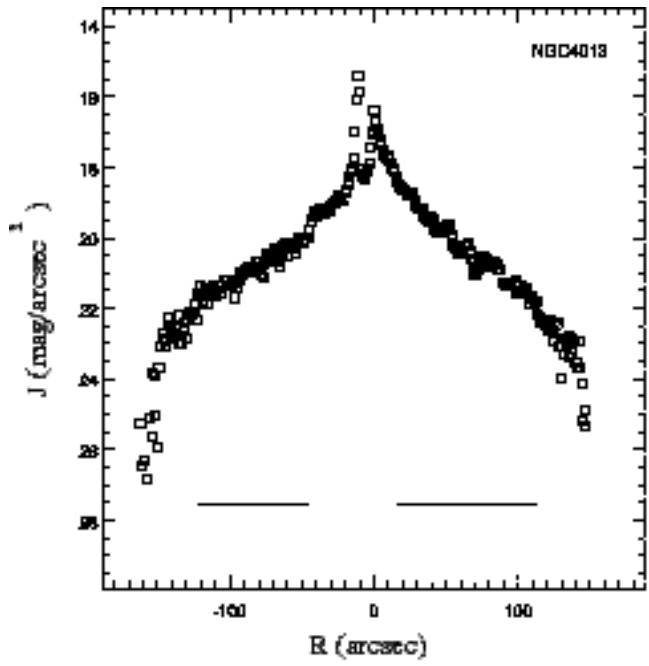
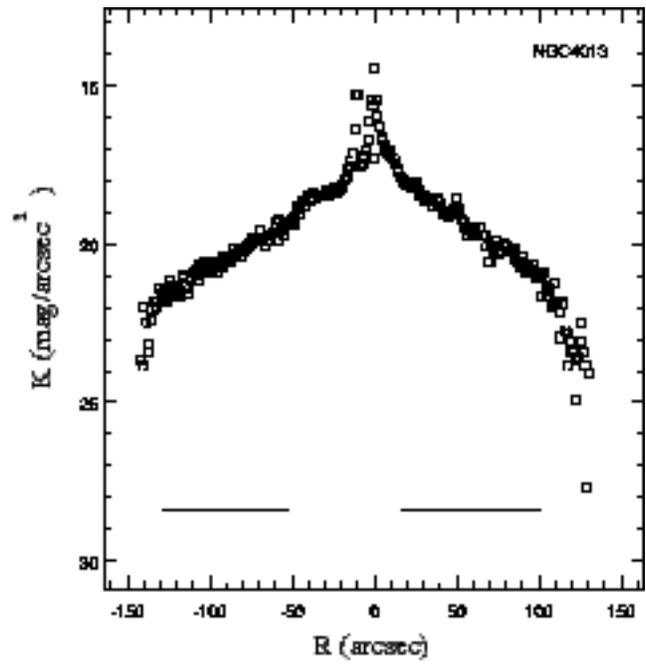
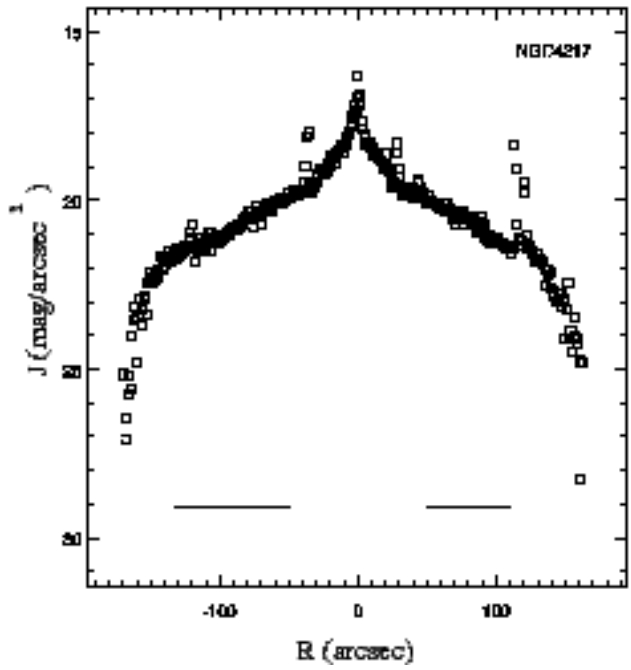
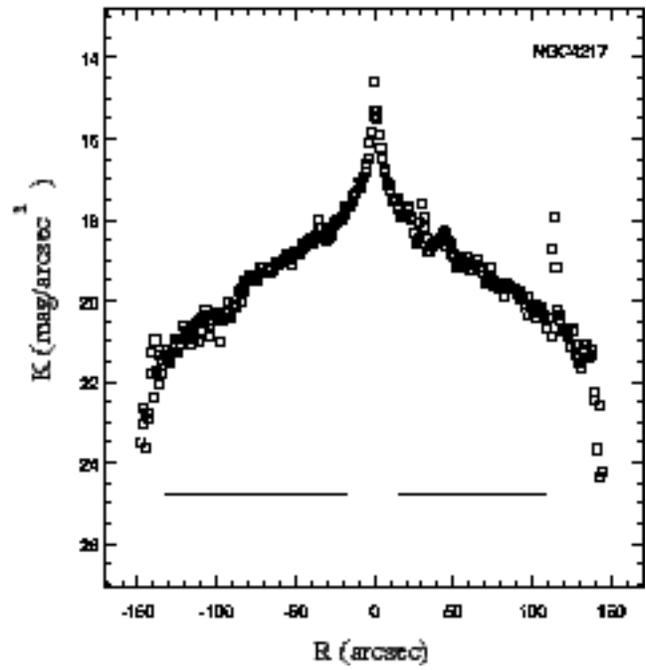
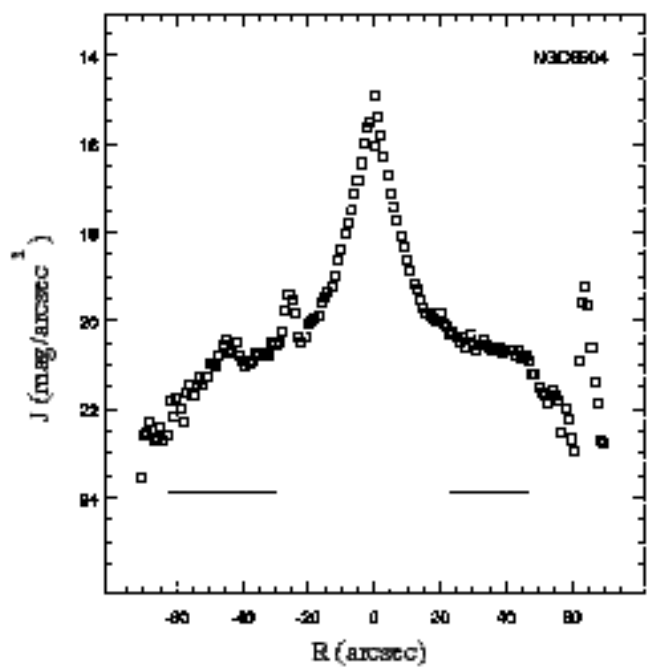
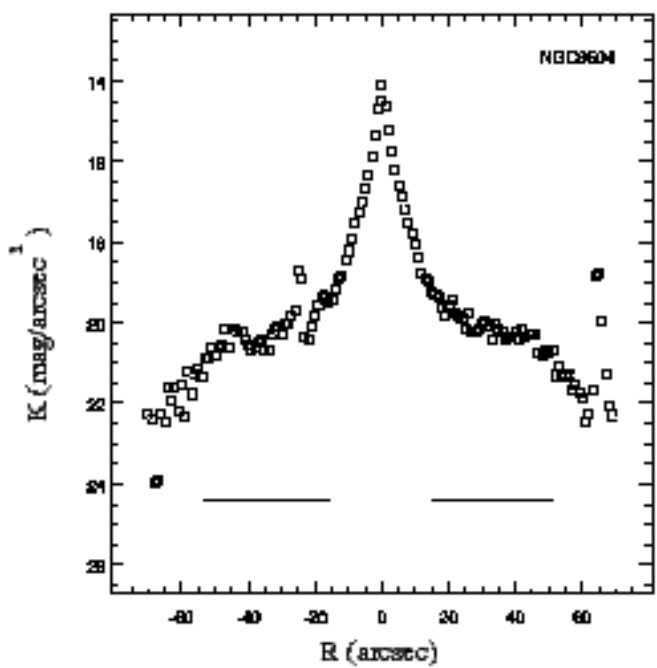
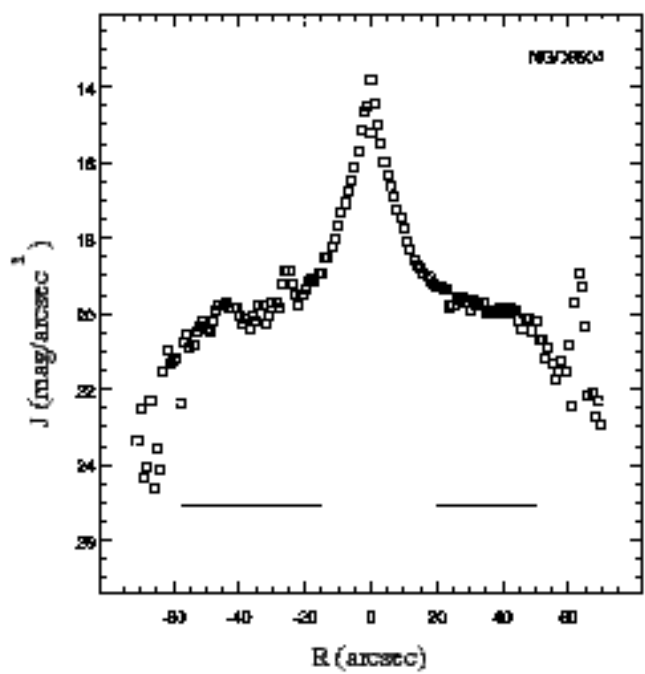
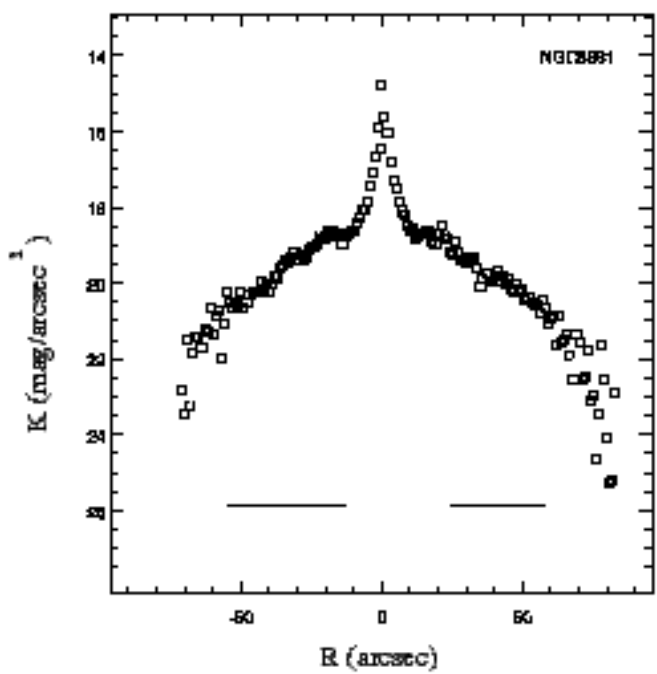
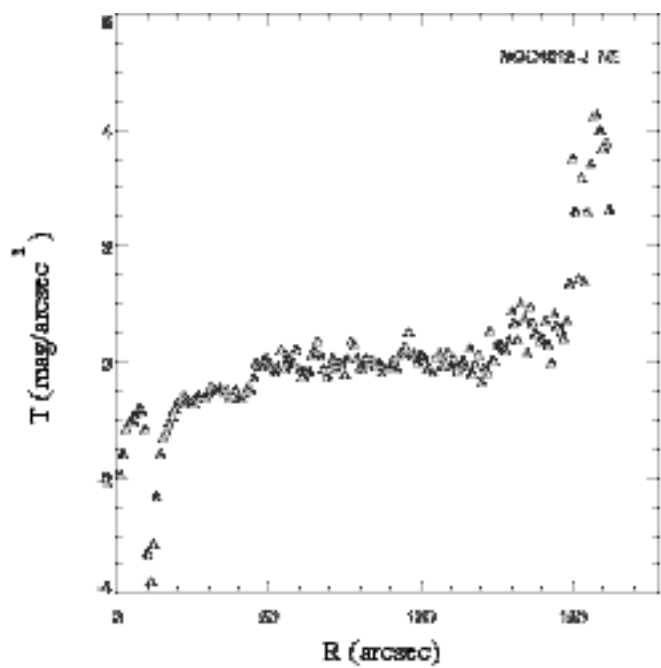
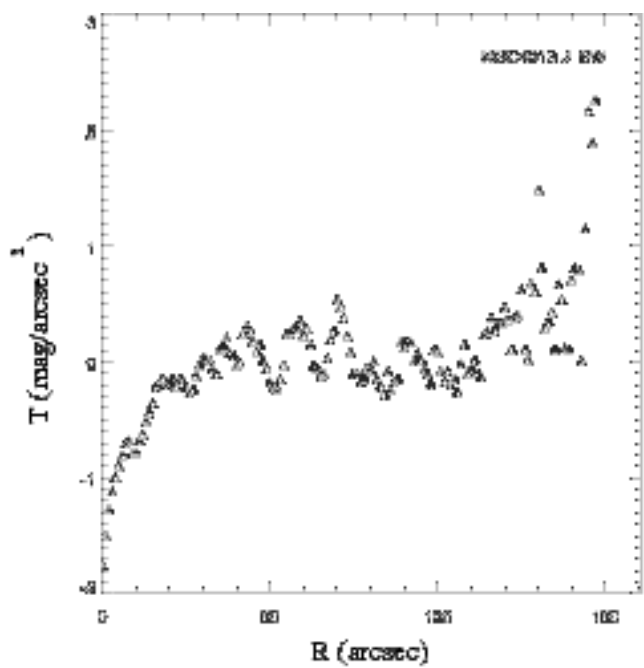
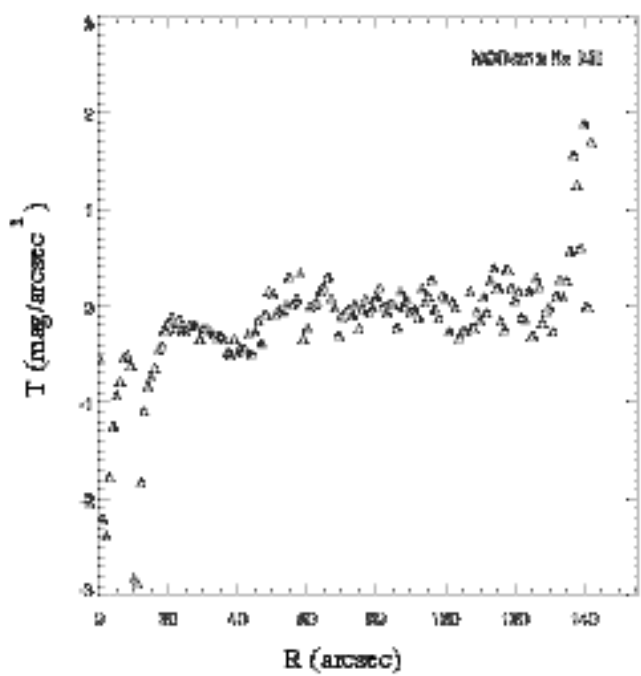
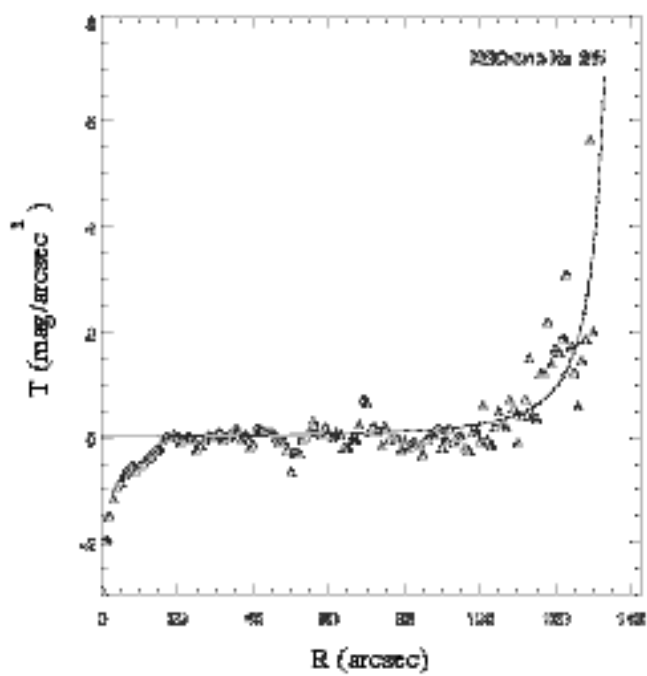
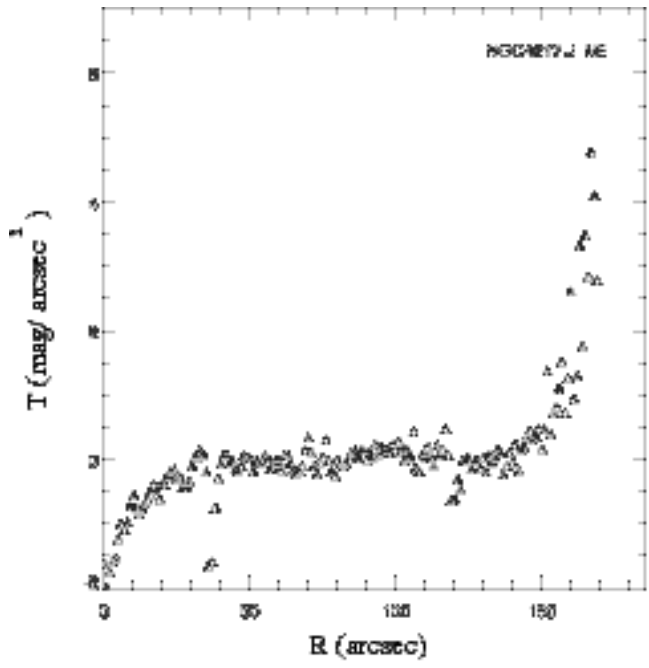
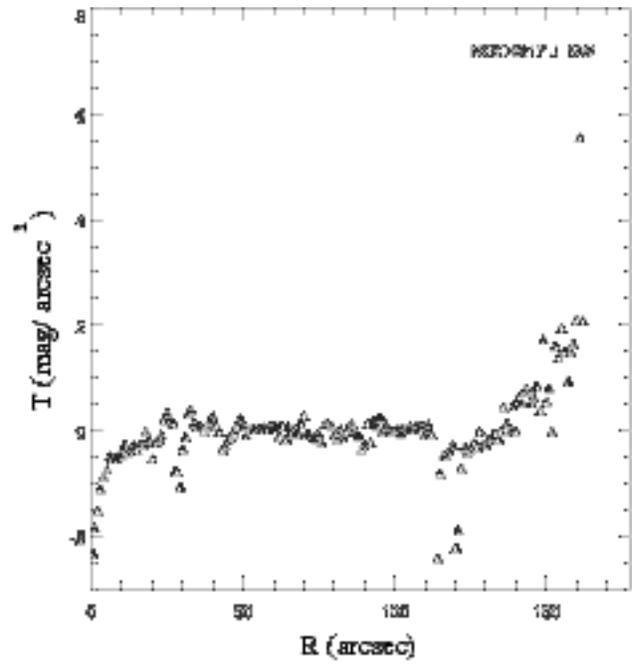
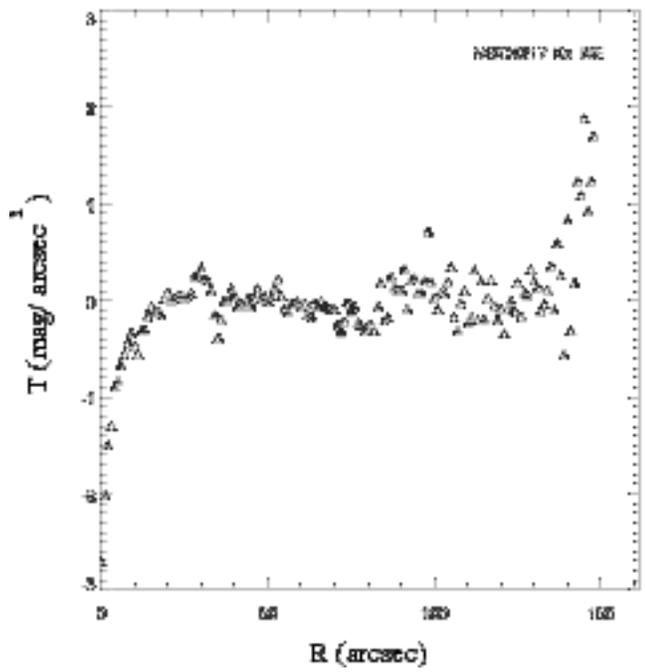
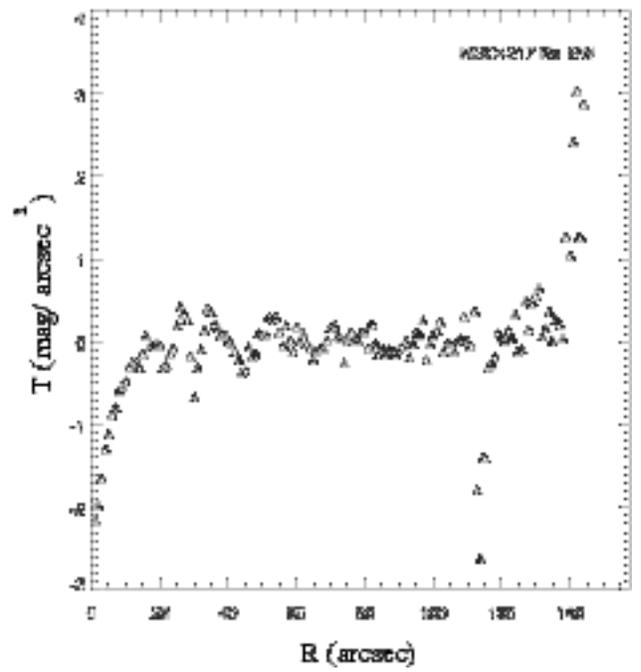
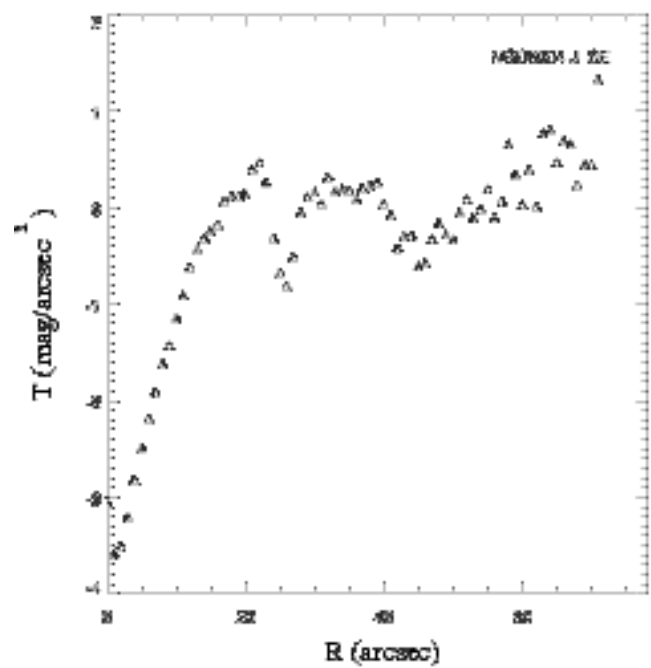
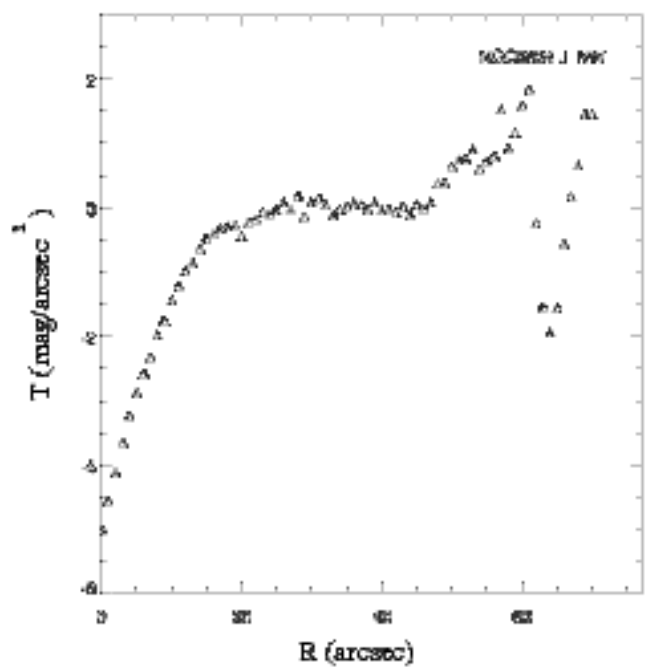
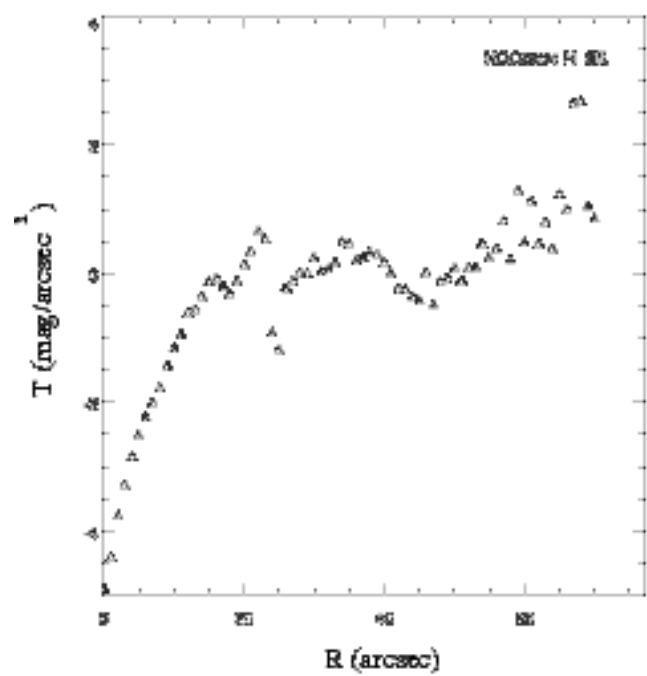
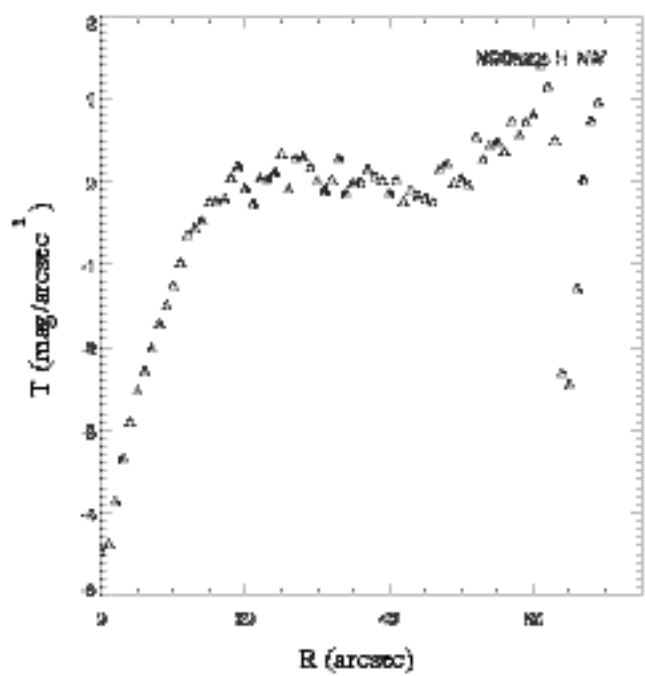
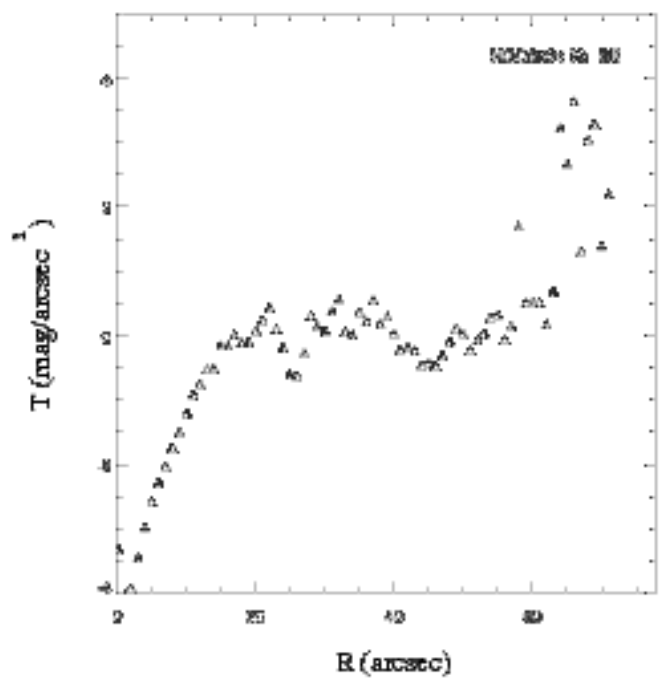
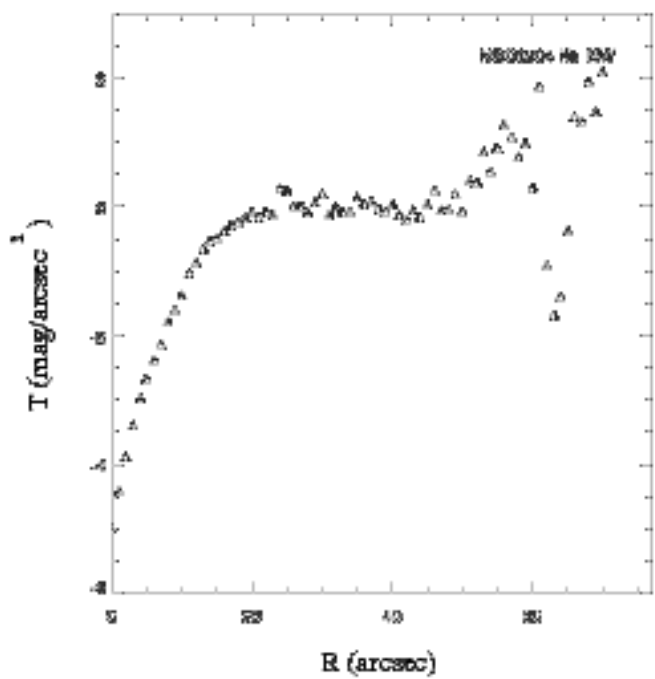
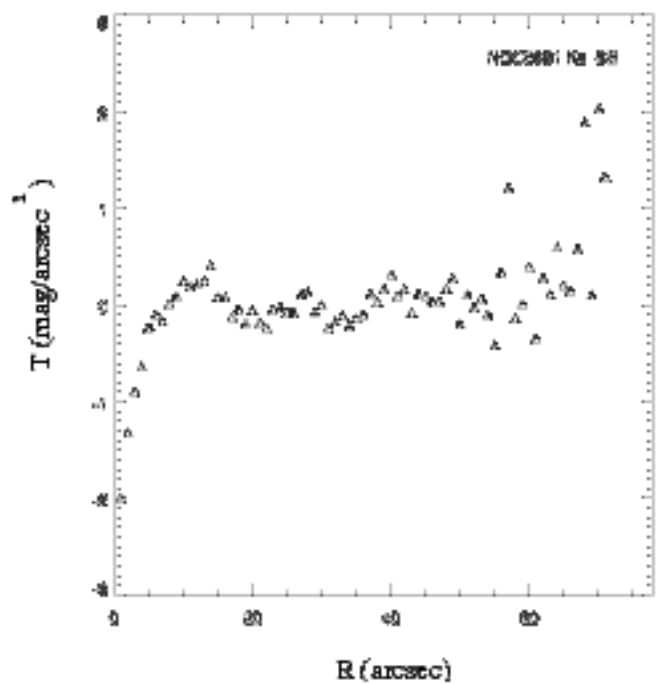
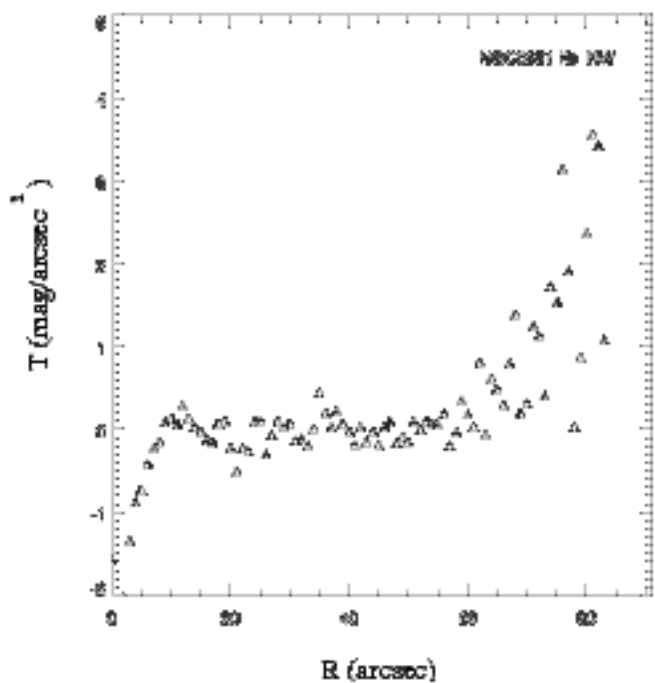
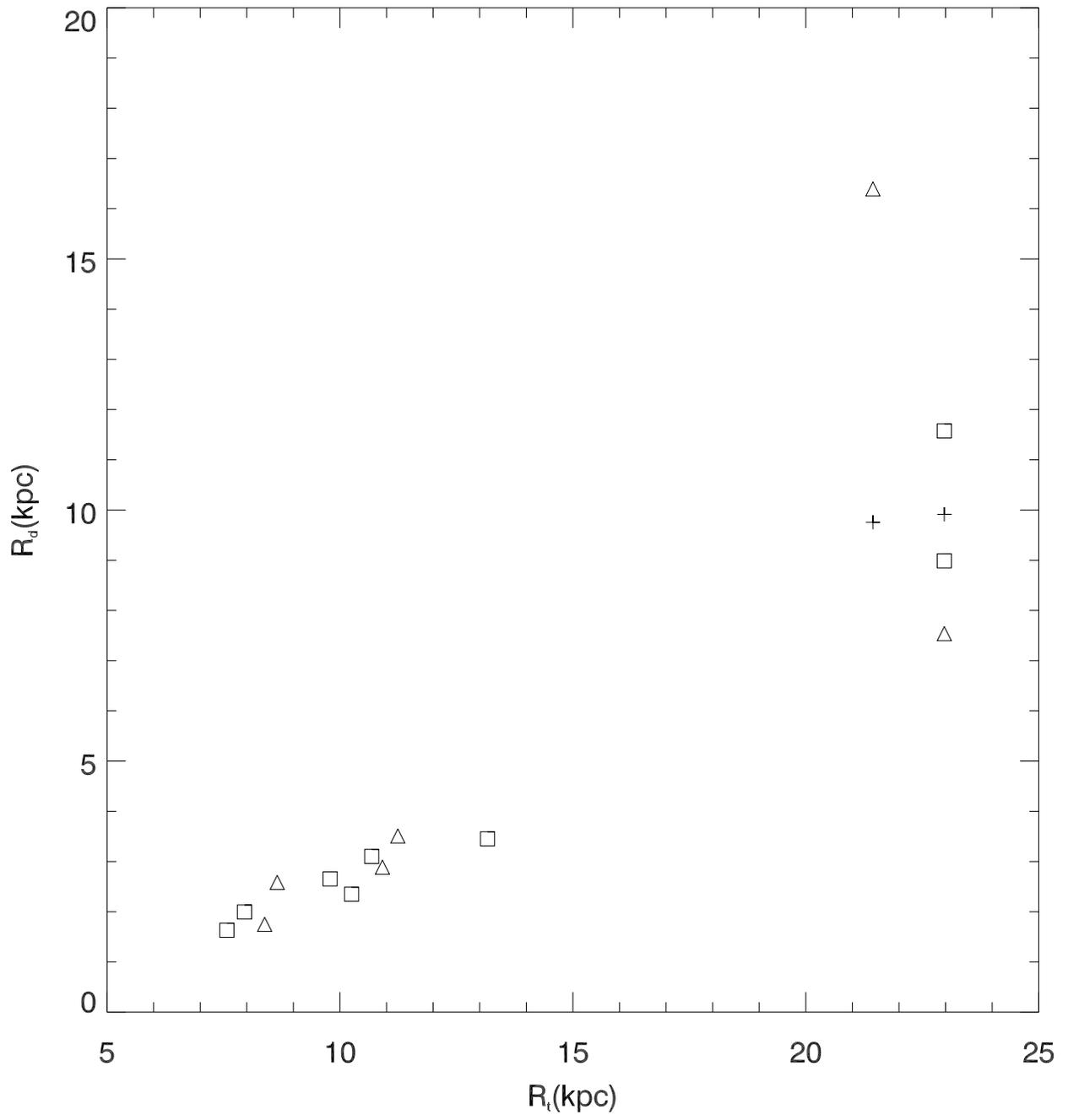
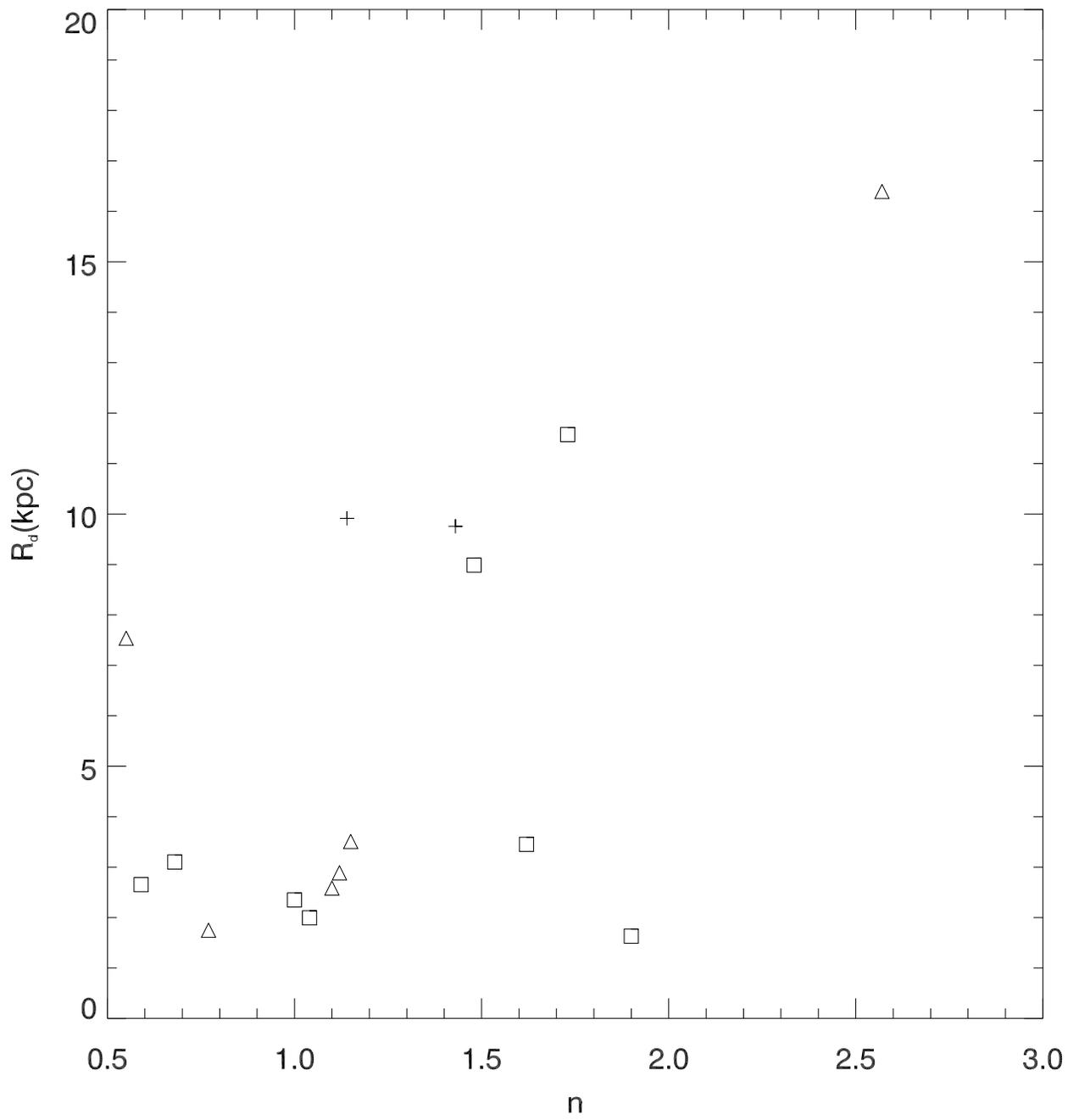
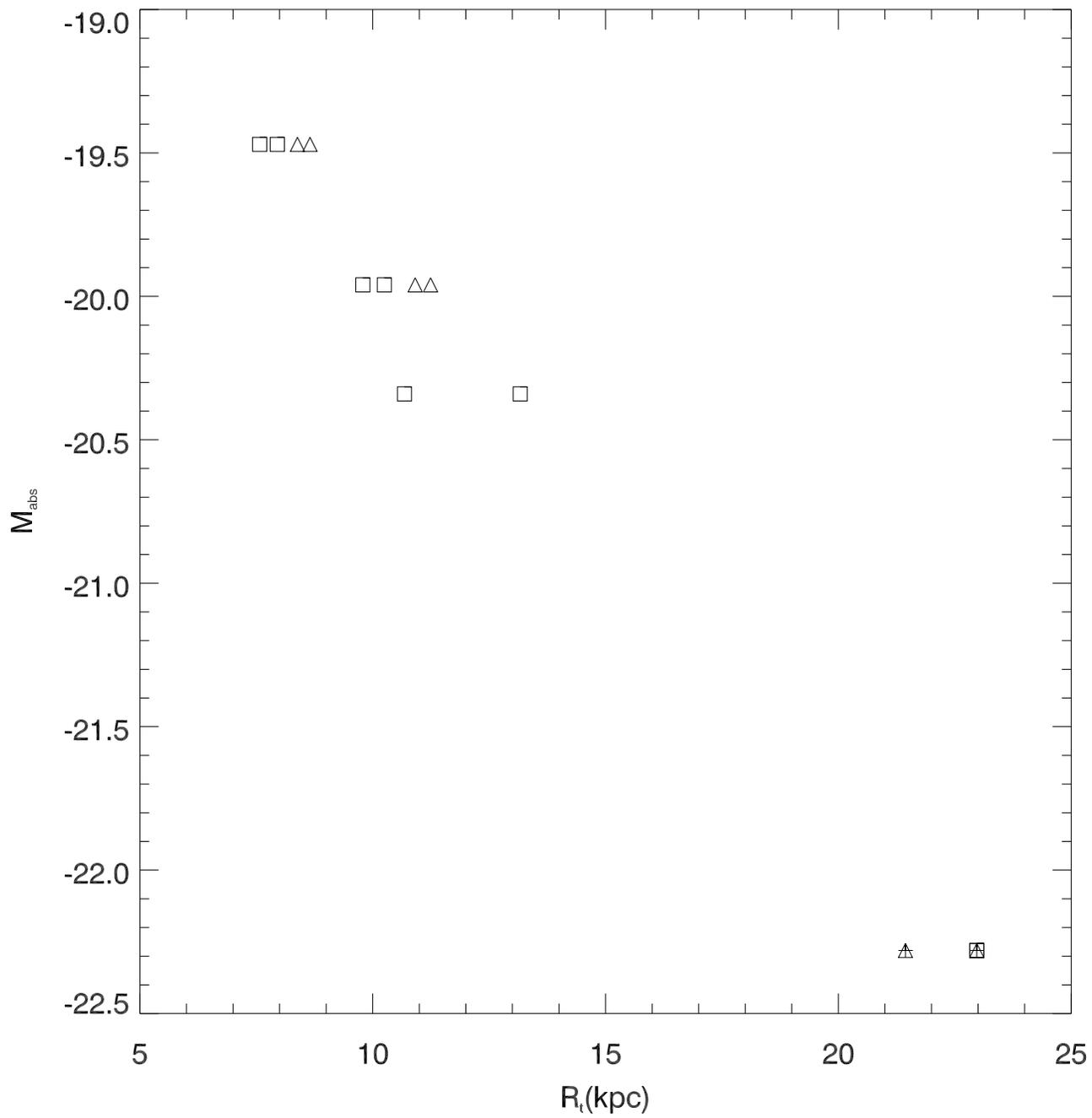
\end{document}